# Implementing a Nordic-Baltic Federated Health Data Network: a case report

Taridzo Chomutare, PhD1,2; Aleksandar Babic, PhD3; Laura-Maria Peltonen, PhD4,5; Silja Elunurm, LLM11; Peter Lundberg, PhD7,8; Arne Jönsson, PhD12; Emma Eneling, BSc7; Ciprian-Virgil Gerstenberger, MSc1; Troels Siggaard, MSc9; Raivo Kolde, PhD6; Oskar Jerdhaf, BSc7; Martin Hansson, BSc10; Alexandra Makhlysheva, MSc1; Miroslav Muzny, PhD1; Erik Ylipää, BSc8; Søren Brunak, PhD9; Hercules Dalianis, PhD10

1. Health Data Analytics, Norwegian Centre for E-health Research, Tromsø, Norway

2. Department of Computer Science, UiT The Arctic University of Norway, Tromsø, Norway

3. Group Research and Development, DNV, 1322 Høvik, Norway

4. Department of Health and Social Management, University of Eastern Finland and Wellbeing services county of North Savo, Kuopio, Finland

5. Department of Nursing Science, University of Turku and Research Services, Wellbeing Services County of Southwest, Turku, Finland

6. Institute of Computer Science, University of Tartu, Tartu, Estonia

7. Department of Radiation Physics, Department of Medical and Health Sciences, Linköping University, Linköping, Sweden

8. Center for Medicine Imaging and Visualization Science (CMIV), Linköping University, Linköping, Sweden

9. *Novo Nordisk Foundation Center for Protein Research, Faculty of Health and Medical Sciences, University of Copenhagen, Copenhagen, Denmark*

10. *Department of Computer and Systems Sciences (DSV), Stockholm University, Stockholm, Sweden*

11. *Institute of Family Medicine and Public Health, University of Tartu, Tartu, Estonia*

12. *Department of Computer and Information Science, Linköping University, Linköping, Sweden*

Corresponding author: Taridzo Chomutare, Norwegian Centre for E-health Research, Sykehusvegen 23, 9019 Tromsø, Norway, taridzo.chomutare@ehealthresearch.no, +47 47680032

## ABSTRACT

**Background:** Centralized collection and processing of healthcare data across national borders pose significant challenges, including privacy concerns, data heterogeneity and legal barriers. To address some of these challenges, we formed an interdisciplinary consortium to develop a federated health data network, comprised of six institutions across five countries, to facilitate Nordic-Baltic cooperation on secondary use of health data. The objective of this report is to offer early insights into our experiences developing this network. **Methods:** We used a mixed-method approach, combining both experimental design and implementation science to evaluate the factors affecting the implementation of our network. **Results:** Technically, our experiments indicate that the network functions without significant performance degradation compared to centralized simulation. **Conclusion:**

While use of interdisciplinary approaches holds a potential to solve challenges associated with establishing such collaborative networks, our findings turn the spotlight on the uncertain regulatory landscape playing catch up and the significant operational costs.

[Case report: total words 1999, 2 tables, 1 figure]

# INTRODUCTION

The advent of deep learning (DL) models, especially Large Language Models (LLMs) has ushered in a new era of Artificial Intelligence (AI) with profound implications for various domains, including healthcare. However, model training needs vast amounts of data, often as centralized repositories, which raises significant concerns regarding data privacy, security, and heterogeneity. To address these challenges, decentralized approaches such as federated learning (FL) have emerged as promising alternatives, where models can be trained securely across multiple institutions without sharing patient data. In this study, we report on a Nordic-Baltic federated health data network initiative designed to facilitate the development of DL models for healthcare applications.

Most research to date has focused on controlled laboratory settings. Though valuable, these studies often lack the complexity and unpredictability of real-world applications. Our work represents one of the first attempts to implement an FL network in a practical, operational environment across national borders. This report's objectives are to show technical feasibility and offer actionable insights into the practical implications and broader contextual factors.

## MATERIALS AND METHODS

To achieve these two objectives, we consider a mixed-method approach that combines two complementary fields of research; data science and implementation science. This interdisciplinary approach to studying translational research in healthcare exploits methods native to each of the fields, resulting in a unique blend of both qualitative and quantitative analysis.

**Federated network requirements and experimental design**

The consortium (**FederatedHealth**) includes Norwegian Centre for E-health Research (NSE) at the University Hospital of North Norway, DNV's Healthcare Program (DNV, Norway), Stockholm University (DSV, Sweden), County Council of Östergötland/University Hospital of Linköping (CCÖ, Sweden), University of Copenhagen (KU, Denmark), University of Turku (UTU, Finland), University of Tartu (UT, Estonia), Cambio (Sweden) and Omilon (Denmark). The datasets available in this consortium are electronic health record (EHR) data from large regional hospitals in the respective countries. However, at this stage we use non-sensitive, publicly available[1], health data for a named entity recognition task; extracting keywords related to medical findings in clinical texts.

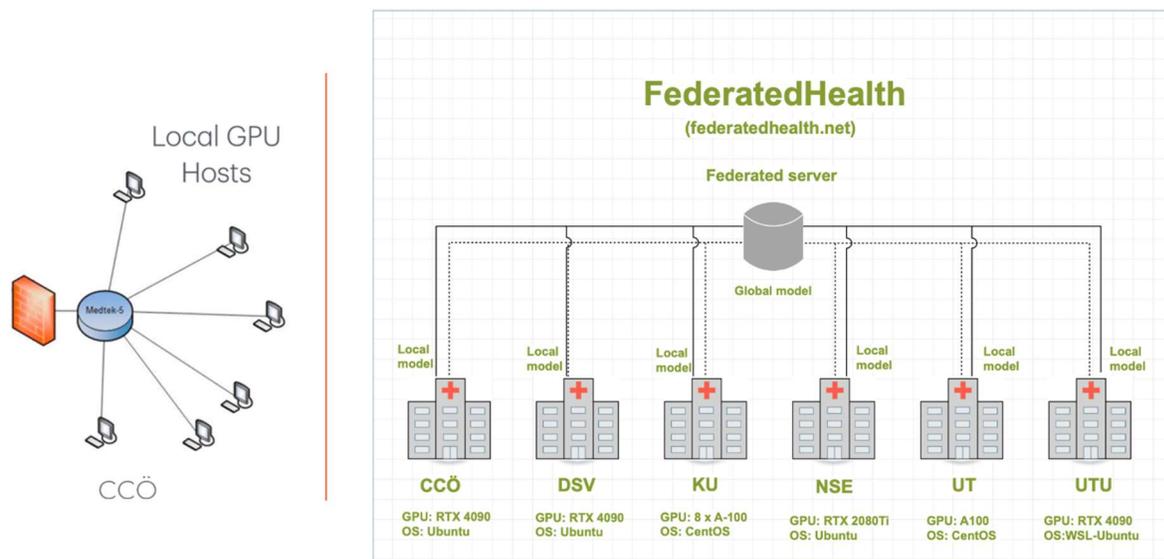

*Figure1:* *The federated network consists of 6 sites with EHR data, two of which are actual hospital sites (NSE, CCÖ), and the rest are universities with close ties to their regional hospitals, an EHR vendor and a health tech business. Example of a single node or site, consisting of six workstations (CCÖ) on Ubuntu (left insert), where sensitive medical data are stored locally, in a separate storage network folder.*

The network illustrated in Fig1 shows the participating institutions and their hardware and software. The central coordinating server is responsible for model aggregation while each institution trains the model locally using its own data, and only the model updates (e.g., gradients, weights) are shared. The primary goal of these basic experiments is to demonstrate that the key technical aspects function as expected, before going in with sensitive data, as a risk mitigation measure. These experiments assess how the system handles varying network configurations and conditions: (i) training with a different number of clients in the network, and (ii) training with data imbalance among the clients.

**Qualitative analysis through implementation science**

Even though implementation science is a recent field, it has demonstrated its potential to enhance our understanding of AI implementation in healthcare[2]. To complement the technical experiments, we also employed an implementation science framework, *Consolidated Framework for Implementation Research* (CFIR)[3], which is one of the more popular meta-theoretical frameworks that can be applied to different stages of an implementation. Its theoretical constructs are especially suited to reflecting on barriers and facilitators. Therefore, we use this framework to systematically organize our collective experience elicited through interdisciplinary workshops and brainstorming sessions.

## RESULTS

**Performance comparison**

As shown in Table 1, our experimental results demonstrated that the number of clients participating in the FL network had no significant impact on the performance of the global model, suggesting the global model generalized effectively across varying client participation levels.

**Table 1:** *Number of clients and performance after 20 training rounds*

| No. of clients | 2 (simulation) | 2 | 4 | 6 | Roth *et al.* (4 client simulation) |
|---|---|---|---|---|---|
| Precision | 0.97 | 0.96 | 0.95 | 0.95 | 0.96 |
| Recall | 0.98 | 0.97 | 0.97 | 0.97 | 0.97 |
| F1-score | 0.97 | 0.97 | 0.96 | 0.96 | 0.97 |

In our data imbalance experiment (see Table 2), where the clients contributed varying amounts of data to the FL process, we found no significant differences in the performance of the global model. These results generally confirm that our technical setup is sound, but the results should be interpreted with caution since the network is relatively small and consists of six sites. In addition, we did not consider class imbalance and the time factor, since training with a physically distributed setup, with varying network bandwidth and hardware requires a longer training time compared to a well-controlled simulation environment.

*Table 2: Data imbalance in percentages and performance after 20 training rounds*

| Data balance (%) | 50/50 | 75/25 | 90/10 |
|---|---|---|---|
| Precision | 0.96 | 0.96 | 0.96 |
| Recall | 0.97 | 0.97 | 0.97 |
| F1-score | 0.97 | 0.97 | 0.97 |

**Consolidated Framework for Implementation Research (CFIR)**

Across the five domains (I-V below), we paid closer attention only to the most pertinent constructs to our context and added some extra constructs we considered important for federated networks, for a total of seventeen points of reflection.

*I. Innovation domain*

**1. Innovation design and costs:** We used an existing FL framework because of budgetary constraints. Combining the practical insights from interdisciplinary brainstorming sessions with the knowledge gained from the literature, we were able to formulate a comprehensive set of functional and non-functional requirements that informed the choice and configuration of our FL framework.

**2. Innovation source:** Our selected FL framework, Nvidia Flare[4], is developed by a generally reputable vendor and incorporates several defensive mechanisms for ensuring the security and privacy of data during federated learning. The framework aligns well with recently published recommendations (2986-2023 - IEEE Recommended Practice for Privacy and Security for Federated Machine Learning).

**3. Innovation complexity and adaptability:** Since the data remains securely stored within each institution, the framework itself can be swapped out or upgraded without altering the underlying data infrastructure. This modularity allows our consortium to adopt new frameworks or algorithms with minimal disruption.

**4. Common data model (CDM):** Harmonizing data in federated learning tasks is a foundational step that ensures reliability, accuracy, fairness and efficiency of the entire task as well as a technical necessity. In this project, we chose to employ a commonly used solution, namely the Observational Medical Outcomes Partnership (OMOP) Common Data Model (CDM), which is an open community data standard.

**5. Data and Model security:** FL enhances privacy by enabling multiple parties to collaboratively train models without sharing raw data. However, this distributed approach also introduces significant security challenges. Data security in FL is focused on protecting sensitive information from unauthorized access, leakage, and inference attacks. To counter these threats, we employed various defense mechanisms such as multifactor authentication and other mechanisms such as differential privacy (see Appendix 2).

**6. Data de-identification:** To prevent the leakage of personal information, text must be de-identified using sanitization tools before uploading them to the federated clients. Similarly, to reduce disclosure risk, large language models should be built using de-identified and pseudonymized text[5-7]. We used clinical text de-identification tools developed within the consortium[8] .

*II. Outer setting domain*

**7. Local conditions and financing:** There seems to be general political support for cross-border collaborations for digital health initiatives within the European area. This is evidenced by multiple funding opportunities aiming to increase use of AI in healthcare, of which the current initiative is a beneficiary.

**8. Legal and ethical considerations:** As there is currently no European-wide harmonized regulation governing the secondary use of health data in a multi-country federated health data network (FHDN) (since the European Health Data Space regulation is not yet applicable), the applicable legislation remains the GDPR, along with national data laws regulating the access and use of health data for R&D purposes. Consortium members jointly determine the purposes and means of data processing within the FHDN, acting as joint controllers with respect to the network. Each consortium member is responsible for fulfilling all legal obligations in their respective countries, including obtaining the necessary administrative and/or ethical committee permits required, to ensure lawful data handling and to uphold the project's integrity across all jurisdictions. The process addressing ethical and legal considerations should be seen as cyclical, rather than linear, because some issues should be addressed repeatedly during the process for ethical and legal risk minimization.

*III. Inner setting domain*

**9. Available resources:** Collectively, our consortium has all the expertise we need (data science, IT, cybersecurity, ethics and legal, as well as healthcare) but there are large disparities at each site in the different countries, presenting unique challenges.

**10. Funding:** We need the different expertise represented in each country or site, and the cost implications of such a setup are substantial. Our current consortium's financial resources are not nearly enough, underscoring the need for strategic resource allocation to ensure long-term viability.

**11. Materials and equipment:** In addition to personnel costs, there are infrastructural expenses for deploying and maintaining the necessary hardware and software at each site. As a requirement, we collectively decided on at minimum a single graphics processing unit (GPU) at each site, since it is easy to scale.

**12. Communications:** The cost of coordinating and managing collaboration across different countries is high, including time spent in regular communication and problem-solving. In our case, a considerable amount of time was spent problem-solving (eg. firewall issues), especially in the initial phases. Since each site has discretion on how they participate, including adhering to their own security policy, the right expertise is required at each site, and this was not always possible.

**13. Culture - code of conduct**: As highlighted in [9], a federation providing access to sensitive health data should be guided by three primary objectives: maintaining patient trust, ensuring the ethical use of sensitive data, and fostering trustworthiness among its members. In scenarios where data remains decentralized and participants come from diverse origins, cultures, or operate under varying legal frameworks, a code of conduct becomes an essential tool. For these reasons, our consortiums developed and implemented a comprehensive code of conduct (see Appendix 1).

*IV. Individuals domain*

**14. Leaders and facilitators:** Our consortium is comprised of senior professors, junior researchers and business executives, a diversity that creates an environment with high-level leaders, opinion leaders and team members that facilitate implementation. However, there are distinct differences in work ethic and culture among the countries in the consortium, despite all the countries being within northern Europe.

**15. Technical jargon:** For our collaborative network to function, it is important to convey complex ideas effectively across disciplines. Our principal strategy for dealing with the problem of technical jargon within our multidisciplinary teams was organizing cross-disciplinary workshops to build a shared vocabulary, and encourage team members to avoid overly technical language during meetings.

**16. Capability:** Ensuring the security and integrity of the federated learning network requires highly skilled individuals at each participating institution. For example, for every FL job, there has to be someone who thoroughly understands the code that is executed on their local client node.

*V. Implementation process domain*

**17. Teaming:** Teaming was our primary tool for dealing with the complexity of the interdisciplinary approach. Creating specialized teams within the consortium proved essential for addressing the diverse range of tasks required to successfully implement our network. These teams operated collaboratively yet independently, ensuring that each task was approached with the appropriate depth of expertise and focus.

# DISCUSSION

Our experimental work demonstrates that FL can be a viable solution for training LLMs on sensitive health data, opening the door to broader collaboration across national borders. Thus, FL can facilitate international research and innovation by allowing institutions to share insights and build more accurate, globally representative AI models.

Our experiences highlight the critical role of interdisciplinary collaboration and structured organizational processes, and also call attention to the legal and economic perspectives. The interdisciplinary framework developed through this work can serve as a foundation for tackling non-technical challenges.

**Limitations**

We did not address data heterogeneity and standardization challenges in this report. Our consortium has wide variation in terms of the EHR data, its formats and amounts, and work on a CDM is ongoing.

**Future work**

Since we have five languages in our consortium (three Germanic and two Finno-Ugric), we have begun focusing on multilingual aspects of training LLMs. We will also investigate the commercial viability of our solution to maximize its impact on global health.

# CONCLUSION

Our experiences show that cross-border implementation of a collaborative network in the current uncertain regulatory environment, though feasible, requires significant human and technical resources. This presents scalability challenges, especially for smaller institutions with limited infrastructure and expertise. Further research is required to explore more efficient and accessible solutions to foster data-driven healthcare innovation across borders.

**Acknowledgements**

We would like to acknowledge and thank members of the FederatedHealth project who have not been named as co-authors, but who contributed to the workshops and discussions.

**Competing interests**

None.

**Funding**

We are grateful to Nordic Innovation for funding this work, under grant no. 407-7003-P22026. The funder played no part in the preparation of this manuscript.

**Data Availability**

We used a publicly available dataset that was originally made available by Doğan et al. [1], and was pre-processed by Roth et al. [4]: https://www.ncbi.nlm.nih.gov/CBBresearch/Dogan/DISEASE/

# Appendix 1: Code of conduct

**FederatedHealth code of conduct (version 1.0)**

This Code of Conduct outlines the standards and principles expected of all participants in the FederatedHealth network. It is designed to foster an environment of trust, respect, and ethical behavior, ensuring the effective and responsible operation of the federated learning initiatives.

*Commitment to Data Privacy and Security*

1. Participants must ensure compliance with all relevant data protection laws and regulations that apply to their data handling practices. This includes adhering to international and national legal requirements for data privacy and security.

2. Participants must actively maintain the confidentiality, integrity and availability of data. This involves implementing and adhering to effective data privacy and security practices, such as secure data transmission, encryption, and access control, to safeguard data from unauthorized access, modification, or exposure.

*Ethical Standards and FAIR Practices*

1. Participants must commit to established ethical guidelines for trustworthy Artificial Intelligence[1], ensuring that models developed do not perpetuate bias or discrimination.

2. Transparency in model development and outcomes is crucial in our networks. Any potential biases or limitations of models must be openly communicated.

3. Participants must try to provide features explaining the decisions of AI models to users.

*Collaboration and Transparency*

1. Participants must engage in transparent and open communication, fostering a collaborative environment.

2. Sharing of insights and knowledge is encouraged, while respecting the confidentiality of sensitive information.

*Compliance with Regulation*

1. All participants must comply with the legal and regulatory frameworks of their respective jurisdictions.

2. International standards and guidelines should be considered and respected where applicable.

---

[1] Ethics guidelines for trustworthy AI, https://digital-strategy.ec.europa.eu/en/library/ethics-guidelines-trustworthy-ai

*Conflict resolution*
1. Participants agree to resolve conflicts through fair and transparent mechanisms.
2. A defined process in the consortium agreement is in place for addressing grievances or disputes within the network.

*Accountability and Responsibility*
1. Participants are responsible for the accuracy and quality of the data they provide.
2. All parties must take responsibility for the impact of their contributions on the network and its outcomes.

*Cultural and Social Sensitivity*
1. Participants must respect the cultural and social differences within the network.
2. Sensitivity towards diverse viewpoints and practices is essential for harmonious collaboration.

*Continuous Improvement and Learning*
1. Participants should strive for continuous improvement in their practices and contributions to the network.
2. Engagement in learning and adaptation to new challenges and technologies is encouraged.

# Appendix 2: Nvidia Flare privacy and security defense methods

**NVIDIA FLARE privacy and security defense methods**

*Fault recovery methods for non-malicious failure*

Addressing the context of non-malicious failures is crucial for maintaining the robustness of Federated Machine Learning (FML) systems as highlighted in[1]. Non-malicious failures refer to system failures that occur without any intentional malicious activity, often resulting from factors such as network issues, participant device limitations, or data inconsistencies. Within the NVIDIA FLARE framework, several measures can indirectly support handling these failures. For example, the framework includes general robustness features such as reliable messaging, which can effectively overcome network glitches and other disruptions.

In terms of fault recovery techniques, NVIDIA FLARE has a potential to facilitate over-participation strategy[1] where more participants than required are participating in each training round to ensure that training can continue even if some participants are slow or unresponsive.

On the other hand, there is no explicit information supporting the use of Asynchronous Secure Aggregation (SecAgg) within NVIDIA FLARE. SecAgg helps to aggregate model updates securely without requiring all participants to contribute within a fixed timeframe, thus making the system more robust to delays and failures. Again, the absence of mention in the documents leads to a high certainty that this technique is not explicitly part of the framework. That said, the framework supports asynchronous communication through its robust infrastructure, utilizing technologies such as gRPC and MPI to manage various messaging patterns and configurations. The communication system in FLARE, managed via the Communicator layer and local gRPC handlers, allows for flexible and non-blocking message operations, significantly decoupling the implementation layers and enabling seamless asynchronous operations. These features suggest that FLARE's current asynchronous communication capabilities could be reasonably expanded to support Asynchronous Secure Aggregation (SecAgg), leveraging its existing protocols and mechanisms designed for efficient and flexible messaging.

Additionally, the practice of creditability evaluation, assessing and prioritizing data owners based on their historical reliability, could not be identified in the NVIDIA FLARE documentation. This technique aims to enhance the reliability of the training process by identifying and giving preference to reliable participants.

Regarding quality control for training data, the documents do not provide any explicit recommendations or requirements such as setting image resolution standards to ensure data quality. Quality control measures are essential to minimize the impact of noisy or suboptimal data on model performance.

Lastly, in addressing mitigating Non-IID data issues, where data across participants may have different distributions, the documents do not mention any strategies such as using a globally shared dataset with a uniform class distribution to balance participants' local data[1]. These strategies are crucial for ensuring that the model is exposed to a representative dataset, thereby improving convergence and stability. The absence of such strategies in the documentation leads to a high certainty of their non-inclusion in the NVIDIA FLARE framework.

## Defensive methods for data attacks

### Secure multiparty computation

Secure Multiparty Computation (MPC) is a technique that enables multiple parties to jointly compute a function over their inputs while maintaining the privacy of those inputs. This approach allows participants to contribute their data to a computation process without revealing the actual data to each other or any central authority, thereby preserving confidentiality[1].

NVIDIA FLARE incorporates several defensive mechanisms to ensure the security and privacy of data during federated learning. One of the key techniques supported by NVIDIA FLARE is Homomorphic Encryption (HE). This method allows for secure training, as demonstrated in the implementation of Secure XGBoost. In horizontal secure training scenarios, each party encrypts its local histograms before sending them to the federated server for aggregation. The server aggregates the encrypted data and returns the encrypted global histograms to the clients for decryption and further training. In vertical secure training, active parties encrypt the gradients before sharing them with passive parties, ensuring that the gradient information remains protected throughout the process.

Another important mechanism provided by NVIDIA FLARE is the facilitation of Multi-party Private Set Intersection (PSI). This technique is crucial for applications such as secure user-ID matching and feature overlap discovery in vertical learning scenarios. PSI allows parties to identify common data points without disclosing their respective datasets, thereby maintaining data privacy.

Additionally, NVIDIA FLARE includes a secure computation framework comprising a comprehensive set of tools and plugins designed to manage secure computation between parties. This framework ensures that data remains protected through encryption and other privacy-preserving methods throughout the federated learning process, reinforcing the overall security of the computation.

### Differential privacy

Differential Privacy (DP) is a mathematical framework designed to protect individual entries within a dataset. The foundational principle of DP is that any algorithm is deemed differentially private if the analysis of its output does not reveal whether any specific individual's data was included in the input dataset. In the realm of Federated Machine Learning (FML), DP techniques typically involve the addition of noise or the use of randomized responses to the data or to the information exchanged during the learning process. This added noise effectively masks the contributions of individual data points, thereby preserving privacy while maintaining the potential for meaningful learning.

In the context of NVIDIA FLARE, Differential Privacy is implemented through various methods, notably through the use of the "SVTPrivacy" filter. This filter facilitates differential privacy by systematically applying noise during the federated learning process, thereby safeguarding individual data points. Specifically, DP techniques within FML in NVIDIA FLARE involve the application of filters to data as it is transmitted between parties, effectively obscuring individual contributions by adding noise.

### AI-based approaches

After a thorough examination related to AI-based defensive mechanisms as indicated in [1], the relevant findings indicate that NVIDIA FLARE does not directly implement, for example, advanced label handling methods specifically akin to the label disguise technique discussed. Additionally, there are no findings pointing to the employment of autoencoders for label transformation or controlled confusion as proposed in[1]. That said, the framework is open source and potentially can provide a flexible and secure base upon which such methods could be built and implemented.

### Other methods

In Federated Machine Learning (FML), privacy-preserving methods such as Secure Multiparty Computation (MPC) and Differential Privacy (DP) are often associated with significant computational overhead. This can be particularly challenging for AI devices with limited resources. To address this issue, a native privacy-preserving approach has been proposed, which involves evaluating the Privacy-Risk-Level of the AI device and collecting user-specific privacy preferences. These factors are then used to estimate the Privacy-Overhead, guiding the selection of an optimal combination of privacy-preserving methods tailored to the device's capabilities and the user's privacy requirements. This strategy is validated through user consent before implementation, ensuring a balance between robust privacy protection and computational efficiency.

NVIDIA FLARE's architecture is designed to support robust privacy and security mechanisms through its framework and configurable filters. Although it does not explicitly detail every step as recommended by some references, it provides a flexible and secure foundation upon which such methods could be built and implemented. NVIDIA FLARE facilitates personalized privacy settings by allowing the application of site-specific privacy policies. Each site can configure its own privacy settings through local privacy policy files, which guide the selection and implementation of appropriate privacy-preserving methods.

While privacy risk evaluation and personalized security methods or levels are not explicitly mentioned in the available documentation, NVIDIA FLARE's architecture supports robust security policies via its security framework. This framework covers identity and data privacy protection measures, indicating potential groundwork for implementing such risk evaluation mechanisms. However, the system does not explicitly detail mechanisms for collecting individual user privacy preferences. Instead, privacy settings and policies appear to be predefined by administrators or researchers, rather than being dynamically collected from individual end-users.

Additionally, there are no details provided on a quantitative approach to estimate Privacy-Overhead based on Privacy-Risk-Level and Privacy-Preference within the NVIDIA FLARE system. The documentation discusses the implementation and use of privacy-preserving filters but does not delve into assessing their computational or operational overheads. Moreover, user consent in the NVIDIA FLARE system is primarily managed through predefined site policies and administrative control, rather than through personalized consent mechanisms. The security framework includes mechanisms for authentication, authorization, and auditing to ensure that operations are approved by relevant authorities, aligning with robust security practices.

*Defensive methods for model attacks*

### Adversarial training

Adversarial training is a defensive technique in machine learning aimed at enhancing the robustness of models against adversarial attacks. These attacks involve making small, often imperceptible changes to input data that can cause a machine learning model to make significant errors. This is particularly important in the context of Federated Learning (FML), where multiple participants may not be fully trusted, making adversarial training crucial for strengthening the resilience of the global model.

Although the NVIDIA FLARE GitHub repository does not explicitly mention support for adversarial training, the framework provides several tools and functionalities that can be adapted to implement this technique. The NVIDIA FLARE Client API allows for easy adaptation of centralized training code to a federated learning setting. By modifying local training loops, adversarial sample generation and training on these samples can be integrated into the workflow. Specifically, the Client API can be used to initialize the environment, receive the global model, perform adversarial training, and send the updated model back to the server.

NVIDIA FLARE also supports advanced algorithms and workflows, such as FedAvg, FedProx, and SCAFFOLD, which can be customized to implement adversarial training techniques. By injecting adversarial samples into the local training process, the robustness of the model can be improved. The Controller APIs and workflow mechanisms, like Scatter and Gather, provide flexibility to introduce adversarial training steps during the local training processes on the clients' side.

The framework allows for the definition of custom Python classes and methods that generate adversarial samples and include them in the training loop. By following the patterns provided in existing examples, custom components and configurations can be created for specific training workflows, enabling the introduction of adversarial training into the federated learning process.

NVIDIA FLARE's support for integration with popular deep learning frameworks like PyTorch and TensorFlow further facilitates the implementation of adversarial training. These integrations allow for the creation and training of models with adversarial samples generated using techniques from libraries such as Foolbox or the Adversarial Robustness Toolbox (ART), alongside the standard training code.

Lastly, NVIDIA FLARE's cross-site validation workflows enable the evaluation of model robustness across multiple sites. These mechanisms can be extended to assess the model's resilience to adversarial attacks during both the training and validation phases, ensuring that the global model maintains its integrity against potential adversarial threats.

## Malicious participants detection

Detecting malicious participants in Federated Machine Learning (FML) is a crucial defensive strategy to protect the global model from poisoning attacks. These attacks occur when participants, whether colluding or acting independently, intentionally submit manipulated or malicious updates—such as gradients or weights—with the aim of degrading or compromising the performance of the global model.

NVIDIA FLARE supports several defensive mechanisms designed to detect and mitigate the impact of malicious participants in FML. One such mechanism is the use of robust aggregation methods, including the coordinate-wise median and geometric median. These methods help to identify and exclude outliers or abnormal updates from participants during the aggregation process, thereby protecting the integrity of the global model. By incorporating these techniques, NVIDIA FLARE helps ensure that malicious updates are detected and mitigated effectively.

In addition to robust aggregation, NVIDIA FLARE supports privacy-preserving algorithms such as Differential Privacy (DP) and Homomorphic Encryption (HE). These algorithms assist in pre-training evaluations and ongoing data inspections, which are critical for scrutinizing the data before and during the training processes. By identifying abnormal updates early, these privacy filters play a significant role in detecting potential malicious activities.

NVIDIA FLARE also includes an auditing mechanism that enhances transparency and accountability by logging significant events within the system. This mechanism allows for the tracking and logging of updates from participants, making it easier to identify and isolate patterns that may indicate malicious behavior. The auditing feature contributes to the overall security by providing a detailed record of participant activities during the training process.

Another important feature in NVIDIA FLARE is the detection of unsafe components. The system includes a custom checker that inspects and validates components before they are used in job execution. If a component is found to be unsafe, such as one that might leak sensitive information, an "UnsafeComponentError" can be raised, preventing the compromised component from affecting the federated learning process. This feature adds an additional layer of protection against potential threats.

Finally, NVIDIA FLARE integrates strict security policies, including the use of TLS certificates for secure provisioning and communication among participants. The system also offers an event-based security plug-in capability that allows for localized authentication and authorization. This added security measure helps in vetting participants and their updates during federated learning sessions, further safeguarding the global model from malicious activities.

## Resilient aggregation

In the context of Federated Machine Learning (FML), resilient aggregation is a method designed to tolerate Byzantine failures, which refers to the challenge of managing participants in the FML system that might behave unpredictably or maliciously. Such failures can occur during the aggregation of parameter vectors, such as gradients or weights, that are uploaded by participants. Ensuring the resilience of this aggregation process is crucial for maintaining the integrity of the global model.

According to information from the NVIDIA FLARE GitHub repository, the framework supports defensive mechanisms that align with resilient aggregation methods, as outlined in the IEEE 2986-2023 (Recommended Practice for Privacy and Security for Federated Machine Learning) standard. NVIDIA FLARE includes mechanisms to handle Byzantine fault tolerance, treating the aggregation of updates from clients with strategies to mitigate the influence of potentially faulty or malicious updates. Federated learning algorithms and workflows, such as FedAvg, can be configured to incorporate robustness techniques specifically designed to handle Byzantine faults. While NVIDIA FLARE does not explicitly reference resilient aggregation techniques like FLTrust, it includes existing federated learning algorithms and privacy-preserving techniques that help mitigate the risk of poisoning attacks.

Although there is no explicit mention of trust score calculation using ReLU-clipped cosine similarity, NVIDIA FLARE's secure and flexible design allows for the implementation of such techniques if needed. The framework's existing privacy-preserving filters and frameworks, such as Differential Privacy and Homomorphic Encryption, could be components in calculating and utilizing trust scores, further enhancing the resilience of the aggregation process.

NVIDIA FLARE also supports weighted parameter aggregation, which allows the contributions of each client to be weighted differently. This feature is essential for implementing aggregation mechanisms that reduce the impact of abnormal updates from malicious participants. Additionally, the framework supports normalization steps on received updates to prevent participants from sending disproportionate updates, ensuring a fair contribution from all clients during aggregation. These features align with the goals of resilient aggregation by promoting balanced and secure contributions from all participants.

Algorithmic support within NVIDIA FLARE includes a range of federated learning algorithms, some of which intrinsically handle issues related to resilience against abnormal updates and participant failures. Examples include FedProx and SCAFFOLD, which adjust for server-client deviations and employ correction terms to ensure robust and stable model convergence, even in the presence of faulty participants.

Moreover, NVIDIA FLARE's comprehensive security framework supports many aspects of secure federated learning operations, including identity security, communication security, and data privacy protection. These elements indirectly contribute to the resilience of the aggregation process. The framework's auditing mechanism further enhances robustness by providing transparency and accountability for operations carried out during federated learning sessions.

While NVIDIA FLARE does not explicitly reference resilient aggregation techniques, it has the infrastructure, flexibility, and mechanisms to support such techniques. This makes the implementation of resilient aggregation feasible within the capabilities of NVIDIA FLARE, ensuring a robust defense against Byzantine failures and malicious activities in federated learning environments.

### Digital signature for model

Digital signatures play a crucial role in protecting the intellectual property rights (IPR) of models developed through Federated Learning (FL) processes. In FL, where multiple participants collaborate to build a global model, there is a heightened risk that the model could be copied, misused, or redistributed without proper authorization. Digital signatures provide a mechanism to verify the legitimate ownership of these models, ensuring they are used according to the rights of the owners.

NVIDIA FLARE supports several defensive mechanisms to embed and verify digital signatures within models, thereby safeguarding intellectual property. One key mechanism is the incorporation of tamper-proof techniques that embed signatures within models during the training process. This is essential for establishing the legitimacy and ownership of the model. For example, the inclusion of signatures in generated configurations during provisioning reflects the system's capacity to embed verifiable signatures within models, ensuring their authenticity.

NVIDIA FLARE also employs TLS certificates for secure provisioning, which is vital for the identity verification of all communicating parties in the federated learning process. The provisioning tool helps create a startup kit for each participant, including tamper-proof mechanisms like digital signatures. These signatures ensure the authenticity and integrity of the federated model during both communication and deployment.

Moreover, NVIDIA FLARE supports event-based security plug-ins that enhance job-level function authorizations. These functionalities allow for the authentication and authorization of models deployed during federated learning operations through the use of digital signatures. This method ensures that signatures are both embedded and verified in a manner that protects the model's intellectual property rights.

Identity security and authorization are further emphasized within NVIDIA FLARE's security framework, which uses Public Key Infrastructure (PKI) technology. Combined with digital signatures, this technology ensures that identities involved in federated learning are trusted and verified, protecting models from unauthorized access and usage.

Additionally, NVIDIA FLARE's auditing mechanism logs significant events and actions, such as model training and deployment tasks. These logs include crucial metadata that can help track the ownership and integrity of models through embedded digital signatures. By maintaining detailed records, this mechanism ensures transparency and accountability in the federated learning process, reinforcing the protection of intellectual property.

Reference: [1] IEEE 2986-2023: IEEE Recommended Practice for Privacy and Security for Federated Machine Learning